\documentclass[twocolumn,showpacs,preprintnumbers,amsmath,amssymb]{revtex4}
\usepackage{graphicx}
\usepackage{dcolumn}
\usepackage{bm}

\begin{document}

\title{Local field effect as a function of pulse duration}

\author{Denis V. Novitsky}
\email{dvnovitsky@tut.by}
\affiliation{%
B.I. Stepanov Institute of Physics, National Academy of Sciences of
Belarus, \\ Nezavisimosti~Avenue~68, 220072 Minsk, Belarus.
}%

\date{\today}

\begin{abstract}
In this note we give semiclassical consideration of the role of
pulse duration in observation of local field effects in the regime
of optical switching. We show that the main parameter governing
local field influence is the ratio of peak Rabi frequency
corresponding to medium inversion and Lorentz frequency of the
medium. To obtain significant local field effect, this parameter
should be near unity that is valid only for long enough pulses. We
also discuss the role of relaxation and pulse shape in this
processes.
\end{abstract}


\maketitle

The concept of local field was introduced in the second half of the
nineteenth century by Hendrik Antoon Lorentz and Ludvig Valentin
Lorenz \cite{Lor}. They demonstrated that the microscopic (local)
electric field $\bf{E}_L$ acting on atoms or molecules of the medium
is different from the macroscopic applied field $\bf{E}$. This
difference is due to polarization of the medium $\bf{P}$ and
describes the near dipole-dipole (NDD) interactions between atoms or
molecules. The well-known expression for the local field in the case
of isotropic homogeneous media is
\begin{equation}\label{Lor_cor}
{\bf E}_L = {\bf E}+\frac{4\pi}{3} {\bf P}.
\end{equation}
Utilization of this relation leads to the classic Clausius-Mossotti
equation between microscopic (molecular polarizability) and
macroscopic (dielectric permittivity) parameters of the medium
\cite{Jackson}. Local field correction (\ref{Lor_cor}) is a good
approximation in the case of nonresonant dense gases, liquids and
solids. Moreover, it can be used to determine refractive index even
for such quantum medium as Bose-Einstein condensate \cite{Morice,
Ruost1, Ruost2} including the effects of atomic correlations.

It turned out that it leads to some fundamental effects if one
considers radiation interacting with a dense collection of resonant
two-level atoms. This system known as a dense resonant medium should
contain many atoms within a cubic resonant wavelength \cite{Bowd93}.
The strength of NDD interactions between atoms of this medium is
measured by value of the Lorentz frequency $\omega_L=4 \pi \mu^2 C /
3 \hbar$, where $\mu$ is a transition dipole moment, $C$ is atom
concentration per unit volume, $\hbar$ is the Planck constant. The
most studied effect induced by presence of local field is the
intrinsic optical bistability which results in two-valued dependence
of population difference between ground and excited states on light
intensity in stationary regime. This effect was predicted
theoretically \cite{Hopf} and then observed experimentally
\cite{Hehlen}. The condition of bistability existence can be
formulated as an inequality $b=\omega_L T_2 > 4$ \cite{Friedberg},
where $T_2$ is the transverse relaxation time. Realistic estimates
of value of $b$ show that, as a rule, it does not exceed several
units, i.e. $b \leq 10$. For example, for gaseous media with typical
parameters $\mu^2=10^{-38}$ erg cm$^3$, $T_2=10^{-9}$ s$^{-1}$,
$C=10^{20}$ cm$^{-3}$ \cite{Afan98} we have $b\approx4$. Condensed
matter, e.g. excitonic media, possesses substantially greater dipole
moments, but they are compensated by relatively small atomic
concentrations \cite{Afan02}.

Local field correction results in some remarkable effects on pulse
propagation in resonant medium. Some of them are connected with new
solitary wave properties, such as distinctions of soliton form from
standard hyperbolic-secant envelope and its area from $2 \pi$
\cite{Bowd91}. NDD interactions play crucial role in generation of
the so-called "incoherent" solitons \cite{Afan02}. They influence
soliton formation in the so-called resonantly absorbing Bragg
reflectors \cite{Cheng}, pattern formation in lasers \cite{Calderon}
and ultrashort, few-cycle pulse propagation in dense resonant medium
\cite{Xia, Xie}. The authors of Ref. \cite{Cren92} considered
ultrafast optical switching of the medium between ground and excited
states due to action of a coherent pulse, i.e. a pulse which
duration is much less than relaxation times of the medium,
$t_p<<T_1, T_2$. Switching was obtained for the pulses with peak
Rabi frequencies $\Omega_p=\mu E_p/\hbar$ ($E_p$ is the peak
amplitude of electric field) approximately equal to Lorentz
frequency of the medium, i.e. $\Omega_p / \omega_L \approx 1$. This
is valid independently of pulse area, however, if the pulse is very
short, it contains only a small fraction of $\pi$ and, obviously,
cannot excite the medium. Therefore, this switching effect holds
true only for pulses long enough, namely, for $\omega_L t_p>1$. This
condition can be rewritten as $b t_p/T_2>1$ and, taking into account
pulse coherence, gives $b>>1$, that seems not to be realistic. In
the next paper \cite{Scalora}, more moderate condition was
considered, $\omega_L t_p \leq 1$, together with taking into account
propagation effects. The results of that work were obtained for
pulses of picosecond durations. On the other hand, in femtosecond
regime, the influence of NDD interactions on pulse propagation was
reported to be negligible, at least for realistic values of $b$ and
$\omega_L$ \cite{Novit}.

In this note we carefully examine the role of pulse duration in
appearance of local field effects. We assume the pulse to be
intensive enough to excite the medium, i.e. the regime of ultrafast
switching is considered. Therefore the main dimensionless parameter
of our research is $\psi=\Omega_0 / \omega_L$, where $\Omega_0$ is
the characteristic peak Rabi frequency corresponding to the pulse
that switches the medium. $\Omega_0$ can be found due to the
conception of pulse area. Indeed, if we take the equality
\begin{eqnarray}
2\frac{\mu}{\hbar} \int^\infty_{-\infty} Edt = 2 \pi \label{area}
\end{eqnarray}
and assume the pulse to have Gaussian shape
$E=E_0\exp(-t^2/2t_p^2)$, we obtain
\begin{eqnarray}
\Omega_0=\sqrt{\frac{\pi}{2}} \frac{1}{t_p}. \label{om0}
\end{eqnarray}
So, the main parameter is
\begin{eqnarray}
\psi=\frac{\Omega_0}{\omega_L}=\sqrt{\frac{\pi}{2}}
\frac{1}{\omega_L t_p}. \label{psi}
\end{eqnarray}
This value allows us to say whether local field correction is
significant or not. It is seen that $\psi$ is dependent on pulse
duration.

Our main thesis is that local field effects can be observed when the
ratio $\psi$ is near unity (relatively long pulses), while they can
be neglected in the case $\psi>>1$ (short pulses). Further we prove
this statement directly by numerical simulations of pulse
propagation inside a dense two-level medium. The model used is based
on the semiclassical Maxwell-Bloch system for population difference
$W$, microscopic polarization $R$, and electric field amplitude
$\Omega'=\Omega/\omega=(\mu/\hbar\omega)E$ (in dimensionless form)
\cite{Bowd93, Cren96, Novit},
\begin{eqnarray}
\frac{dR}{d\tau}&=& i \Omega' W + i R (\delta+\epsilon W) - \gamma_2 R, \label{dPdtau} \\
\frac{dW}{d\tau}&=&2 i (\Omega'^* R - R^* \Omega') -
\gamma_1 (W-1), \label{dNdtau} \\
\frac{\partial^2 \Omega'}{\partial \xi^2}&-& \frac{\partial^2
\Omega'}{\partial \tau^2}+2 i \frac{\partial \Omega'}{\partial
\xi}+2 i \frac{\partial \Omega'}{\partial
\tau} \nonumber \\
&&=3 \epsilon \left(\frac{\partial^2 R}{\partial \tau^2}-2 i
\frac{\partial R}{\partial \tau}-R\right), \label{Maxdl}
\end{eqnarray}
where $\tau=\omega t$ and $\xi=kz$ are dimensionless arguments;
$\delta=\Delta\omega/\omega$ is the normalized detuning of the field
carrier (central) frequency $\omega$ from atomic resonance;
$\gamma_1=(\omega T_1)^{-1}$ and $\gamma_2=(\omega T_2)^{-1}$ are
the rates of longitudinal and transverse relaxation, respectively;
$\epsilon=\omega_L/ \omega$ is the normalized Lorentz frequency;
$k=\omega/c$ is the wavenumber, and $c$ is the light speed in
vacuum. Here we assume that the background dielectric permittivity
of the medium is unity (two-level atoms in vacuum). Equations
(\ref{dPdtau})-(\ref{dNdtau}) are derived in the framework of the
rotating wave approximation (RWA) which requires $\Omega'<<1$
\cite{Allen}. This condition is satisfied throughout the paper. In
Eq. (\ref{Maxdl}) we do not use slowly-varying envelope
approximation (SVEA) which cannot hold true even for thin films of
the medium as noted in Ref. \cite{Scalora}. Description based on
Eqs. (\ref{dPdtau})-(\ref{dNdtau}) does not take into account such
processes as multiple scattering, radiation reabsorption and
spontaneous emission which result in quantum corrections of
Lorentz-Lorenz relation \cite{Fleis1, Fleis2}. However, many usual
effects of light propagation such as self-induced transparency can
be correctly treated in semiclassical approximation \cite{Allen}.

In our calculations we use Gaussian pulses with peak amplitudes
(\ref{om0}) and central wavelength $\lambda=0.5$ $\mu$m. We consider
the case of strict resonance, i.e. $\delta=0$. Initially (before
pulse incidence) the medium is in the ground state, i.e. $W=1$,
$R=0$. Thickness of the layer of the medium is $L=5 \lambda$. NDD
interactions between two-levels atoms provide Lorentz frequency
$\omega_L=10^{11}$ s (note, that $\omega_L<<\omega$). This value is
believed to be high enough according to the typical parameters
above.

\begin{figure*}[t!]
\includegraphics[scale=0.9, clip=]{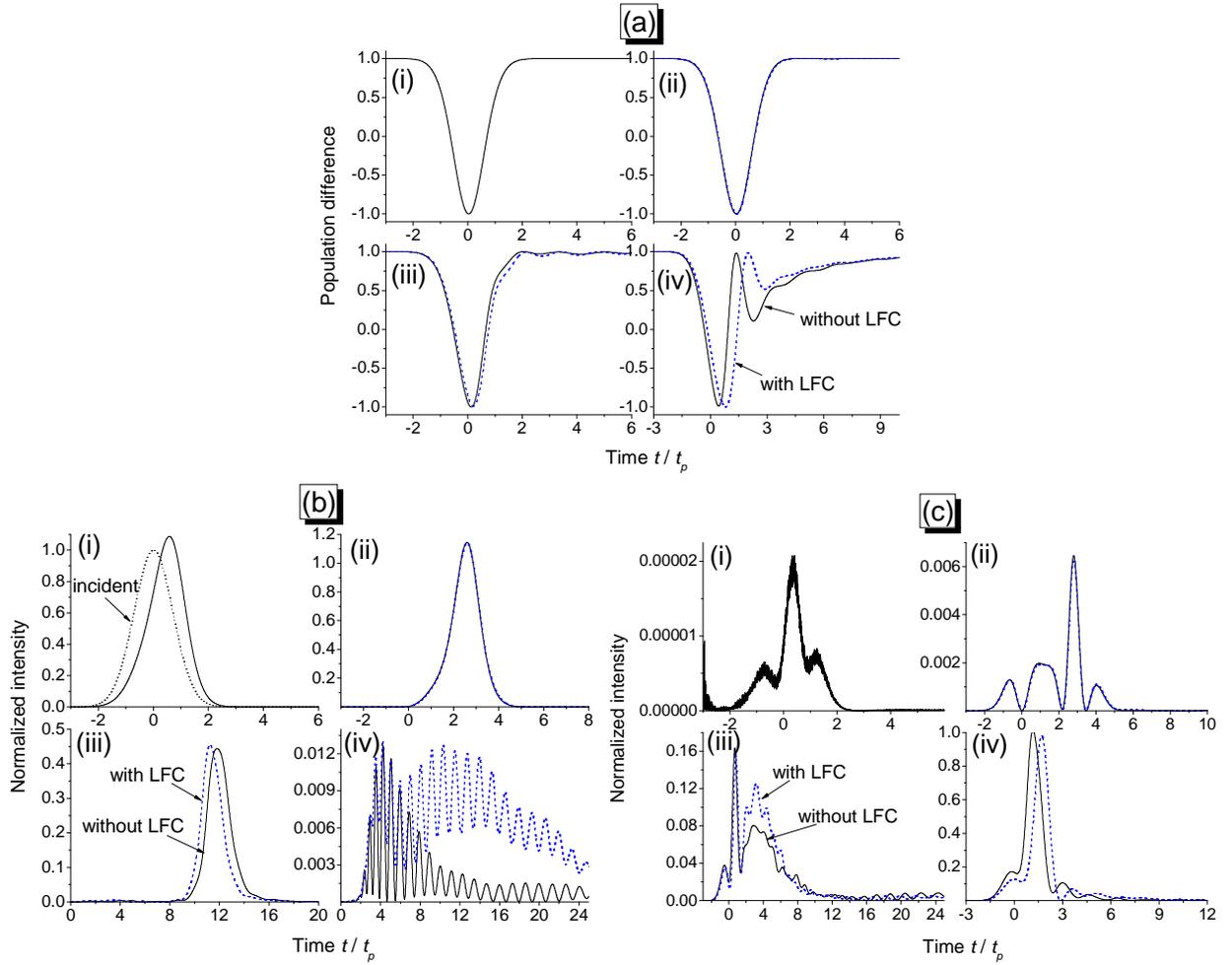}
\caption{\label{fig1} (Color online) (a) Population difference on
the entrance of the medium, (b) transmitted and (c) reflected
radiation at different pulse durations: (i) $t_p=0.1$ ps, (ii)
$t_p=1$ ps, (iii) $t_p=5$ ps, (iv) $t_p=10$ ps. Relaxation is
absent. Results correspond to calculations without local field
correction (LFC) (solid lines) and with it (dashed lines) in Eq.
(\ref{dPdtau}).}
\end{figure*}

First, we consider the case of coherent pulses, i.e. for
phenomenological relaxation terms in Eqs.
(\ref{dPdtau})-(\ref{dNdtau}) we assume $\gamma_1=\gamma_2=0$. This
allows to study pure effect of local field without any side effects
connected with relaxation. As one can see in Figs.
\ref{fig1}(i)-(ii), influence of NDD interactions on dynamics of
pulses with durations $t_p=0.1$ and $1$ ps ($\psi=125$ and $12.5$,
respectively) is negligible. For shorter (femtosecond) pulses this
is valid as well, in accordance with the results of Ref.
\cite{Novit}. Such pulses act as usual $2 \pi$-ones, first inverting
the medium and then returning it exactly into the ground state.
Transmitted pulses demonstrate shape transformation resulting in
pulse compression \cite{Novit}, while reflected radiation is almost
absent. This situation can be treated as self-induced transparency
(SIT) regime.

\begin{figure*}[t!]
\includegraphics[scale=0.95, clip=]{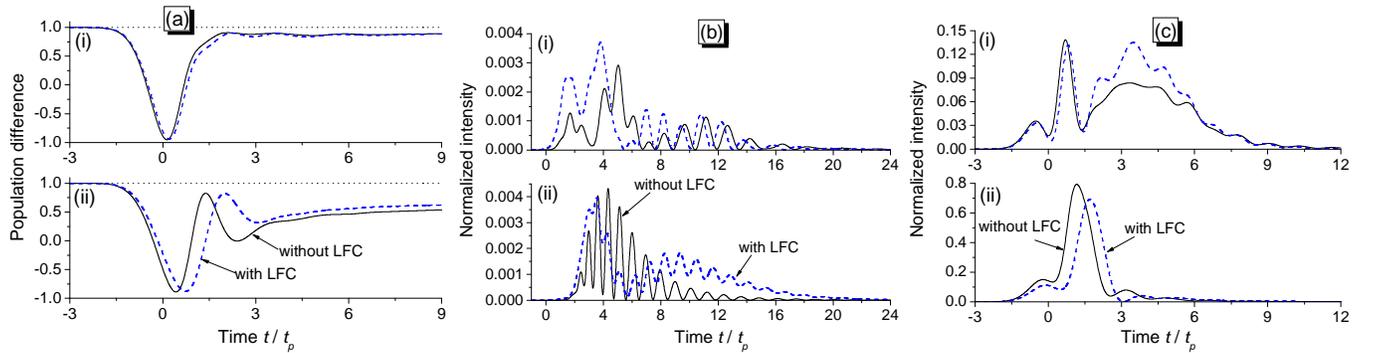}
\caption{\label{fig2} (Color online) (a) Population difference on
the entrance of the medium, (b) transmitted and (c) reflected
radiation at different pulse durations: (i) $t_p=5$ ps, (ii)
$t_p=10$ ps. Relaxation times $T_1=1000$ ps, $T_2=100$ ps. Results
correspond to calculations without local field correction (LFC)
(solid lines) and with it (dashed lines) in Eq. (\ref{dPdtau}).}
\end{figure*}

When we further make pulse duration greater, for $t_p=5$ and $10$ ps
[$\psi=2.5$ and $1.25$, Figs. \ref{fig1}(iii)-(iv)], local field
effects become apparent. Inversion of the medium is reached later in
comparison with the case of absence of NDD interactions. At the same
time, the transmitted pulse is decreasing, while the reflected one
is getting more intensive. At $t_p=10$ ps almost entire initial
energy of radiation is transformed into the reflected pulse.
Perhaps, this is connected with the effect of coherent internal
reflection which was studied in stationary regime earlier
\cite{Malyshev, NovitJOSAB}. However, local field results in larger
transmittance as compared with the case when it is absent [see Fig.
\ref{fig1}b(iv)].

Now let us add phenomenological relaxation. We take typical
parameters $T_1=1000$ ps and $T_2=100$ ps, so that NDD interactions
parameter is $b=\omega_L T_2=1$. For pulse durations $t_p=0.1$ and
$1$ ps the results are almost the same as in relaxation-free case,
see Figs. \ref{fig1}(i)-(ii). But for longer pulses we have to take
into account relaxation. It is seen in Fig. \ref{fig2}a(i) that for
$t_p=5$ ps relaxation results in energy conservation inside the
medium for a long time (population difference does not reach unity)
and, hence, the output (transmitted and reflected) radiation is only
a small fraction of incident one [compare with Fig.
\ref{fig1}(iii)]. For the pulse with $t_p=10$ ps [Fig.
\ref{fig2}a(ii)], relaxation of population difference on the
entrance of the medium is slow, too. However, this results in strong
reflection rather than trapping of pulse energy. The time shift of
both population difference and peak of reflected radiation in the
case of local field correction is seen as well. Therefore, one can
say that local field effects appear in the regime of internal
reflection rather than in the regime of self-induced transparency.

\begin{figure}[t!]
\includegraphics[scale=0.85, clip=]{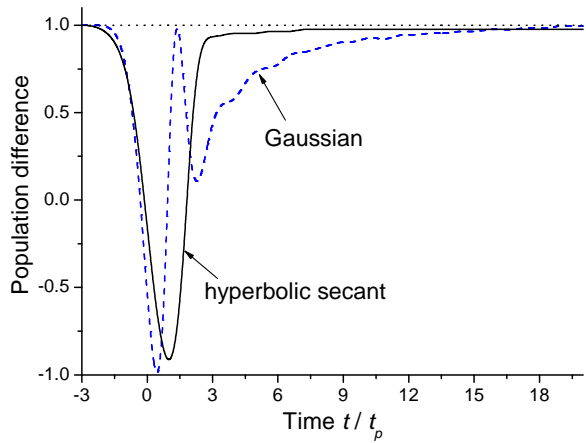}
\caption{\label{fig3} (Color online) Population difference on the
entrance of the medium for different pulse shapes: hyperbolic secant
and Gaussian. Pulse duration $t_p=10$ ps. Relaxation and local field
correction is absent.}
\end{figure}

Finally, we should discuss the question of pulse shape. One can see
in Fig. \ref{fig1}a(iv) that behavior of population difference for
long pulse with $t_p=10$ ps is different from that in Figs.
\ref{fig1}a(i)-(ii) even in the case when local field correction is
absent. This is due to Gaussian shape of such a long pulse. For
comparison we take the invariant pulse with hyperbolic secant shape,
$E=E_0 {\rm sech}(t/t_p)$. The condition (\ref{area}) leads in this
case to the peak Rabi frequency
\begin{eqnarray}
\Omega_0=\frac{1}{t_p}. \label{om0sech}
\end{eqnarray}
Figure \ref{fig3} demonstrates that the curve of population
difference for hyperbolic secant pulse with peak amplitude
(\ref{om0sech}) and duration $t_p=10$ ps is really less deformed as
compared with Gaussian pulse of the same duration. However, all
other peculiarities (e.g. predominant reflection) are still valid in
this case. The same statement is true for qualitative properties of
the effect of local field correction on pulse propagation in the
dense two-level medium considered.

In conclusion, in this note we considered the case of pulse
propagation in a dense two-level medium in the regime of optical
switching. It is clearly demonstrated by direct numerical
calculations that local field effect on pulse propagation in such
media is dependent on pulse duration. The governing parameter $\psi$
is the ratio of peak Rabi frequency (characteristic for medium
switching) and Lorentz frequency of the medium. For short
(femtosecond) pulses this ratio is large, and we have the regime of
self-induced transparency without any significant influence of local
field. In other words, as pulse duration is decreasing, one need to
have much greater Lorentz frequencies (that seems not to be
realistic) to obtain any local field effect. On the other hand, when
Lorentz frequency is increasing as medium is getting more dense, one
has to take into account the processes of multiple scattering (and,
hence, radiation trapping) which was ignored in our study. For long
(picosecond) pulses, such that $\psi \sim 1$, the influence of local
field becomes apparent, while SIT regime transforms into the regime
of coherent internal reflection. On the other hand, the relaxation
processes (just as pulse shape) can be sufficient in the case of
long pulses. The results obtained may be used for proper choice of
the parameters of experiments dealing with local field observation
(at least, in some special experimental geometries).

\end{document}